\documentclass[aps,prd,english,superscriptaddress, longbibliography,11pt,notitlepage]{revtex4}

	\usepackage[colorlinks=true, a4paper=true, pdfstartview=FitV,
linkcolor=blue, citecolor=blue, urlcolor=blue]{hyperref}

\usepackage{amsmath}
\usepackage{amssymb}
\usepackage{graphicx}
\usepackage{hhline}
\usepackage{mwe}
\usepackage{amsbsy}
\usepackage{textcomp}
\usepackage{commath}
\usepackage{subfig}
\usepackage{slashed}
\usepackage{natbib}

\DeclareMathOperator{\sech}{sech}

\usepackage{stackengine}
\stackMath

\makeatletter
\usepackage{babel}
\numberwithin{equation}{section}
\begin{document}
\immediate\write16{<<WARNING: LINEDRAW macros work with emTeX-dvivers
                    and other drivers supporting emTeX \special's
                    (dviscr, dvihplj, dvidot, dvips, dviwin, etc.) >>}

\title{ Fermionic spectral walls in kink collisions }
\author{J. G. F. Campos}
\affiliation{Departamento de Física, Universidade Federal da Paraíba, João Pessoa - PB - 58051-970, Brazil}
\author{A. Mohammadi}
\affiliation{Departamento de Física, Universidade Federal de Pernambuco, Av. Prof. Moraes Rego, 1235, Recife - PE - 50670-901, Brazil}
\author{J. M. Queiruga}
\affiliation{Department of Applied Mathematics, University of Salamanca,
37008, Salamanca, Spain}
\affiliation{Institute of Fundamental Physics and Mathematics, University of Salamanca, 37008 Salamanca, Spain}
\author{A. Wereszczynski}
\affiliation{Institute of Physics,  Jagiellonian University, Lojasiewicza 11, Krak\'{o}w, Poland}
\author{W. J. Zakrzewski}
\affiliation{Department of Mathematical Sciences, University of Durham, Durham DH1 3LE,
United Kingdom}
 
\begin{abstract}
We show that a spectral wall, i.e., an obstacle in the dynamics of a bosonic soliton, which arises due to the transition of a normal mode into the continuum spectrum, exists after coupling the original bosonic model to fermions. This spectral wall can be experienced if the boson or fermion field is in an excited state. Furthermore,  while passing through a spectral wall, an incoming kink-fermion bound state can be separated into purely bosonic kink, which continues to move to spatial infinity and a fermionic cloud that spreads in the region before the wall. 
\end{abstract}

\maketitle

\section{Introduction}
\label{intro}
 The role of internal modes in multi-kink collisions has been widely studied. Now, after more than 40 years of investigations,  we have a reasonably good understanding of the impact of normal and quasi-normal modes on soliton dynamics, especially in the case of bosonic field theories in (1+1) dimensions. 
 
 First of all, such modes may trigger, via the {\it resonant energy transfer mechanism} \cite{Campbell:1983xu}, a chaotic (or even fractal) structure in the final state formation in multi-kink collisions. The most prominent example of such behavior is the kink-antikink scattering in $\phi^4$ theory, where the normal mode of a single kink, called the shape mode, can temporarily store some part of the energy, allowing or not, for the solitons to reappear in the final state \cite{Campbell:1983xu, Sugiyama:1979mi}. 
This has been confirmed only very recently by constructing a collective coordinates model based on two moduli involving the distance between the kinks and the amplitude of the mode \cite{Manton:2021ipk}. Furthermore, the resonant energy transfer can also be switched on by the effective modes existing only in multi-kink configurations, 
such as delocalized modes in antikink-kink collisions in $\phi^6$ theory \cite{Dorey:2011yw, Adam:2022mmm} or by quasi-normal modes \cite{QNM} or even by sphalerons \cite{sphaleron}. In fact, similar chaotic structures have been observed in many other models, see e.g., \cite{Campbell:1986nu, Alonso-Izquierdo:2017gns, Izq-1, Izq-2, Izq-3, Christov:2020zhb, Villa, Tr-1, Kev, MoradiMarjaneh:2022vov, MM}. 
 
 A different phenomenon intimately related to normal modes is the {\it spectral wall} \cite{Adam:2019xuc}. It involves a formation of an arbitrary long-living stationary state due to the transition of a normal mode into the continuum spectrum. In particular, if we scatter a kink with initial velocity $v$ and initially excited mode $\eta$ of amplitude $A$ on another soliton or a non-dynamical background field (impurity), and assume that this mode hits the mass threshold at $a_{sw}$, which can be related to the distance between the antikink and the other kink or impurity,  then this process can proceed through three scenarios: ${\it (i)}$ if $A>A_{cr}$  the kink is 
back-scattered before reaching the point $a_{sw}$ at which the mode enters the continuum; ${\it (ii)}$ if $A<A_{cr}$ then the kink can pass through the point $a_{sw}$ at which the distortion of its motion becomes increasingly weaker as the amplitude decreases; ${\it (iii)}$ and finally, for $A=A_{cr}$, the kink may form a stationary state with its position frozen at $a_{sw}$. Hence, a spectral wall acts as an obstacle in the kinetic motion of a kink. This can be a long-range obstacle as modes may enter the continuum even at a considerable distance before the kink meets other soliton or impurity. Furthermore, this is a very selective obstacle that is experienced only if the pertinent mode is sufficiently excited. Excitation of another mode $\eta'$ does not have any impact on passing the spectral wall by the mode $\eta$. 
 
  Undoubtedly, spectral walls govern low-velocity dynamics of kinks in various solitonic processes, see e.g., antikink-kink collision in the $\phi^6$ model \cite{Adam:2022bus}. In the BPS processes, where a collision occurs via passing through a sequence of energetically equivalent solutions, the position of the spectral wall is uniquely determined by the transition of the given mode to the continuum. Also, it is not affected by details of the initial configuration \cite{Adam:2019xuc, Adam:2021ypm}.
 In non-BPS collisions, a modification arises from a static force between the colliding solitons or soliton-impurity, which changes the position of the stationary solution \cite{Adam:2019uat}. So, the BPS (thin) spectral wall becomes a non-BPS (thick) spectral wall. The main difference is that its position $a_{sw}^{thick}$ does depend on the initial condition, i.e., on the initial velocity of the colliding solitons. However, the selective nature remains unchanged. 
 
A natural extension of kink models in (1+1) dimensions is to couple them to fermions. Due to the rather high complexity of such systems, the back reaction of the fermions on the kinks has usually been neglected. Thus in \cite{Gibbons:2006ge, Saffin:2007ja, Campos:2020iov, Chu:2007xh} the transfer of fermions in kink-antikink (or in general, brane collisions) or 
in kink-impurity was studied, but without back reaction. However, it has only been very recently realized that the back reaction may have a very nontrivial effect on static  \cite{amado2017coupled, Perapechka:2019vqv, Klimashonok:2019iya} as well as on dynamical properties of kinks \cite{campos2022kink, bazeia2022resonance}. 
 In particular, in \cite{campos2022kink}, it was shown that the fermion field generates a force, either attractive or repulsive, depending on how one distributes
 the fermion field on the kink and antikink. Thus, resonance windows in the kink-antikink collisions in the supersymmetric $\phi^4$ model could be highly affected by the presence of the fermion.
What is even more exciting is the fact that the fermion energy could also play a role in the resonant energy exchange mechanism, generating resonance windows for a theory that in a purely bosonic sector is integrable like in the sine-Gordon model \cite{bazeia2022resonance}.

In the present work, we investigate the fate of thin spectral walls in a boson-fermion system in which the back reaction of the fermion on the kink is fully taken into account.
For this, we have chosen the most straightforward nontrivial theory, i.e., a supersymmetric BPS-impurity model.
 In this case, in a purely bosonic version, there is a one-parameter family of energetically equivalent BPS antikink-impurity static solutions representing the soliton at any distance from the impurity. Nonetheless, the spectrum of the linear modes varies as the antikink approaches the impurity. In particular, a normal mode can hit the mass threshold at a certain soliton-impurity distance allowing for the existence of a spectral wall. 

\section{Supersymmetric BPS-impurity model}
\label{Lagrangian}
In this section, we introduce the supersymmetric extension of the BPS-impurity model.  We begin with the following superfield Lagrangian
\begin{equation}
\label{lag:kink}
\mathcal{L}_\Phi=\int d^2\theta \left(\frac{1}{4}D^\alpha \Phi D_\alpha \Phi+W(\Phi)\right),
\end{equation}
with
\begin{eqnarray}
\Phi&=&\phi+\bar{\theta}\psi+\frac{1}{2}\bar{\theta}\theta F,\\
D_\alpha&=&\frac{\partial}{\partial\bar{\theta}_\alpha}-i\left(\gamma^\mu \theta\right)_\alpha\partial_\mu, \\
D^\alpha&\equiv& \bar{D}_\alpha=D_{\beta}\left(\gamma^0\right)_{\beta\alpha},
\end{eqnarray}
and $\bar{\theta}=\theta \gamma^0$, for a minimal $N=1$ SUSY in $1+1$ dimensions. Performing the integration, the Lagrangian takes the following form (see, for example, \cite{Shizuya:2003vm}) 
\begin{equation}
\mathcal{L}_\Phi=\frac{1}{2}\left((\partial_\mu\phi)^2+\bar{\psi}i \slashed{\partial}\psi +F^2\right)+\sqrt{2}W'(\phi)F-\frac{\sqrt{2}}{2}W''(\phi)\bar{\psi}\psi,
\end{equation}
where we have rescaled $W\rightarrow \sqrt{2}W$ and used the normalization $\frac{1}{2}\int d^2\theta \, \bar{\theta}\theta=1$ for the Grassmann integral.

The Lagrangian (\ref{lag:kink}) is invariant under the following $N=1$ SUSY transformations:
\begin{eqnarray}
\label{lag:trans-1}
\delta \phi &=&\bar{\xi}\psi, \\ \label{lag:trans-2}
\delta \psi_\alpha &=&-i(\gamma^\mu\xi)_\alpha\partial_\mu \phi+\xi_\alpha F, \\ \label{lag:trans-3}
\delta F &=& -i\bar{\xi}\gamma^\mu\partial_\mu\psi,
\end{eqnarray}
where $\xi=(\xi_1,\xi_2)$ is a constant Grassmann spinor. The above symmetry leads to the following conserved Noether supercurrent:
\begin{equation}
J_\alpha^\mu=(\gamma^\nu \gamma ^\mu \psi)_\alpha (\partial_\nu \phi)+i\sqrt 2 W(\phi)(\gamma ^\mu \psi)_\alpha .
\end{equation}
The interaction of the field $\phi$ with the non-dynamical background field (impurity) $\sigma$ is taken in a particular form
\begin{equation}
\label{lag:inter}
\mathcal{L}_\sigma=\sqrt{2}\sigma W'(\phi)\left(F-\phi_x\right)-\frac{\sqrt{2}}{2}\bar{\psi}\psi W''(\phi)\sigma,
\end{equation}
which is prescribed by means of a SUSY argument \cite{Adam:2019yst}. 
The interacting part of the Lagrangian (\ref{lag:inter}) changes into a total derivative under the SUSY transformation provided that $\xi^2=0$
\begin{equation}
\label{lag:imp_cur}
\delta \mathcal{L}_\sigma\vert_{\xi^2=0}=\xi^1\partial_t(-i\sqrt{2}\sigma W^\prime(\phi)\psi_2),
\end{equation}
where we have used the following representation $\gamma^0=\sigma_2,\,\, \gamma^1=i\sigma_3$ for the Dirac gamma matrices.
This implies that only one supercurrent, $J^\mu_1$, is conserved, which receives an extra contribution from the supercurrent in (\ref{lag:imp_cur}).
 Consequently, although the impurity explicitly breaks half of the $N=1$ supersymmetry, the other half is preserved, and one BPS sector, with a corresponding Bogomolny equation, survives. 

As usual, the field equation for $F$ is purely algebraic and can be eliminated. For the full model $\mathcal{L}=\mathcal{L}_\Phi+\mathcal{L}_\sigma$ we have
\begin{equation}
\frac{\partial \mathcal{L}}{\partial F}=0\Rightarrow F=-\sqrt{2}W'(\phi)(1+\sigma).
\end{equation}
By inserting this value into the Lagrangian we get
\begin{equation}
 \mathcal{L}^{\text{on-shell}}=\frac{1}{2}\left(\left(\partial_\mu \phi\right)^2+\bar{\psi}i \slashed{\partial}\psi\right)-(1+\sigma)^2 
W'(\phi)^2-\frac{\sqrt{2}}{2}\bar{\psi}\psi W''(\phi)(1+\sigma)-\sqrt{2}\sigma W'(\phi)\phi_x.\label{full-on}
\end{equation}
From now on, we take the semi-classical approach, in which the fermion up and down components are c-numbers. It is easy to understand that 
a collection of a very large number of bosons present coherently in more or
less the same quantum state can be described by a classical field theory. However, this is not the case for the fermions. 
It is possible to prove that it also works for the fermion fields, although
for a different reason and a completely distinct physical interpretation. A detailed demonstration of this assertion can be found in \citep{rajaraman1982solitons}, chapter 9. 

The field equations can then be expressed as 
\begin{equation}
\label{eq:bos}
\phi_{tt}=\phi_{xx}-2 W'(\phi)W''(\phi)(1+\sigma)^2+\sqrt{2}\sigma_x W'(\phi)-\frac{\sqrt{2}}{2}\bar{\psi}\psi  W'''(\phi)(1+\sigma),\\
\end{equation}
for the boson field and
\begin{eqnarray}
\label{lag:fer-1}
i\slashed{\partial}\psi-\sqrt{2}W''(\phi)(1+\sigma)\psi=0,
\end{eqnarray}
for the fermion one. The last term in (\ref{eq:bos}) determines the back-reaction of the fermion on the boson field. The fermions in the supersymmetric model considered here are Majorana. However, taking the semi-classical approach, the back-reaction term for Majorana fermions vanishes. 
For this reason, we shall not restrict ourselves to this case. We take Dirac fermions instead in our analysis. 

Next, we assume that the
 fermion field is initially normalized to unity
\begin{equation}
\label{eq:norm}
\int_{-\infty}^\infty\psi^\dagger(x,0)\psi(x,0)dx=1.
\end{equation} 
Due to the unitarity of the evolution of the Dirac field, the initial normalization is preserved for all times $t$.

Our model possesses a static BPS sector for which solutions obey the field equations (\ref{eq:bos}), (\ref{lag:fer-1}). Specifically, they become:  
\begin{eqnarray}
\phi^s_x&+&\sqrt{2}W'(\phi^s)(1+\sigma)=0, \label{lag:BPS}\\\label{fer-zero}
\psi_0^s&=&\frac{N}{\sqrt 2} 
\begin{pmatrix}
\frac{\phi^s_x}{1+\sigma} \\
0
\end{pmatrix}.
\end{eqnarray}
The first equation can be easily obtained using the SUSY transformations and clearly coincides with the Bogomolny equation of the purely bosonic version of the model. The second equation defines the fermionic zero mode. The existence of the bosonic Bogomolny equation gives rise to a one-parameter family of static, energetically equivalent BPS solutions. 
It should be emphasized that for an arbitrary coupling of the background field to the boson-fermion system, the bosonic Bogomolny equation does not exist. 

Note that solutions of the BPS sector are solutions of the full theory with or without the back-reaction term. Thus, the fermion zero mode does not back-react on the kinks. The multiplicative constant $N$ should be chosen to obey the normalization condition. 
It is well known that there is no back-reaction of the fermionic zero mode on the kinks for the models with energy reflection
 symmetry \cite{jackiw2007chiral, amado2017coupled}.
This is also the case for the Yukawa fermion-kink interaction we are dealing with here. This is consistent with the result we have 
found in (\ref{fer-zero}) as if there is no back-reaction.
 Our result represents a fermionic zero mode attached to the BPS kink. 

Now, let us specify our model. We choose the potential of the $\phi^4$ theory
\begin{equation}
W'(\phi)=m\beta\left[\frac{1-(\phi/\beta)^2}{\sqrt{2}}\right],
\end{equation}
where we have included all the relevant parameters. Next, we perform a rescaling of the fields and coordinates, consisting of $\phi\to\beta\phi$ and $x^\mu\to m^{-1}x^\mu$.
 To maintain the normalization condition \eqref{eq:norm}, the fermionic field is also modified as $\psi\to m^{1/2}\psi$. After these manipulations,
 the bosonic equation of motion becomes
\begin{equation}
\label{eq:bos2}
\phi_{tt}=\phi_{xx}+2 \phi (1-\phi^2)(1+\sigma)^2+\sigma_x(1-\phi^2)+\frac{1}{\beta^2}\bar{\psi}\psi(1+\sigma),\\
\end{equation}
where the parameter $\beta$ controls the strength of the back reaction. The fermionic equation, as well as the normalization condition, remain unchanged. 

Next, we choose the background field in the form
\begin{equation}
\sigma(x)=\alpha \sech^2(x),
\end{equation}
which describes an exponentially localized impurity located at the origin. Here $\alpha$ is a real number. In this case, the BPS solutions take the form:
\begin{equation}
\phi^s(x;c)=-\tanh \left(x+c+\alpha(\tanh(x)+1)\right).
\end{equation}
They describe a $\phi^4$ antikink located at some distance from the impurity. The soliton becomes distorted as it approaches the origin;
 otherwise, its shape coincides with the usual $\tanh$ function. Since all these solutions possess the same energy, there is no static force between 
the antikink and the impurity. Similar behavior also occurs for the fermionic zero mode. 

\begin{figure}
    \centering
    \includegraphics[width=0.65\textwidth]{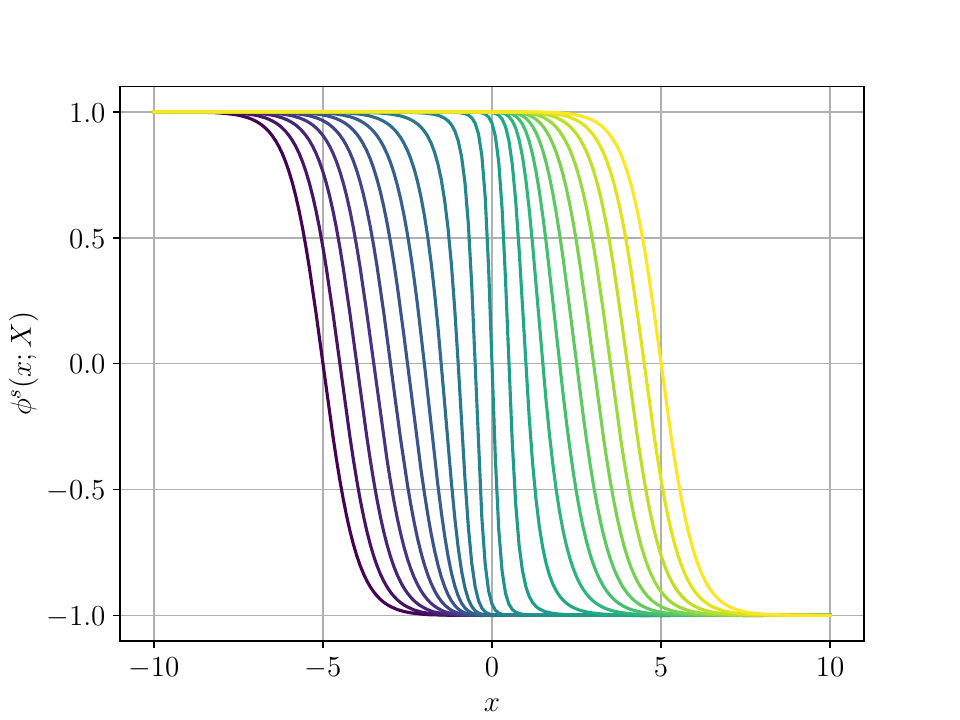}
    \caption{Evolution of the antikink in X space. We set $\alpha=3.0$}
    \label{fig:my_label}
\end{figure}

In our numerical analysis, we can choose an alternative parameterization of the BPS antikink solutions. Namely, $\phi^s(x;X)$, where $X$ is the
 position of its center, which is defined by the point where the scalar field vanishes, that is, $\phi^s(x;X)=0$. For the current model, the 
definition above defines a unique $X$. Fig. (\ref{fig:my_label}) shows the evolution of the above antikink in $X$ space with parameter $\alpha=3.0$. 
Kinks far from the impurity are indistinguishable from the usual $\tanh(x)$ solution.
 Near the impurity, we observe a steeper passage through $\phi=0$.

\section{Spectral structure}
\label{Spectral}
Although the BPS solutions are energetically degenerate, the spectral structure, which is the structure of linear perturbations, changes as
 we alter the value of $c$ (or $X$). Thus, already a linear perturbation of the BPS sector can exhibit fascinating and nontrivial dynamics. Importantly, due to the absence of static force, we can clearly see the role played by the normal modes. 

For simplicity, we will look next at the linear stability equations in the absence of the back-reaction term. To do this, let us consider 
perturbations of the form
\begin{equation}
\phi=\phi^s+ e^{i \omega t} \eta(x).
\end{equation}
Taking (\ref{eq:bos}) into account, we note that $\eta(x)$ satisfies, at the linear level, the following equation:
\begin{equation}
\label{bos:eigen}
\left(-\partial_x^2 + V_ -\right)\eta(x)=\omega^2\eta(x),
\end{equation}
where
\begin{equation}
V_\pm=\mathcal{U}^2\pm\frac{d\mathcal{U}}{dx},
\end{equation}
with
 \begin{equation}
\mathcal{U}=\sqrt{2} W''(\phi^s)\left(1+\sigma\right).
 \end{equation}
  \begin{figure}
    \centering
    \includegraphics[width=0.7\textwidth]{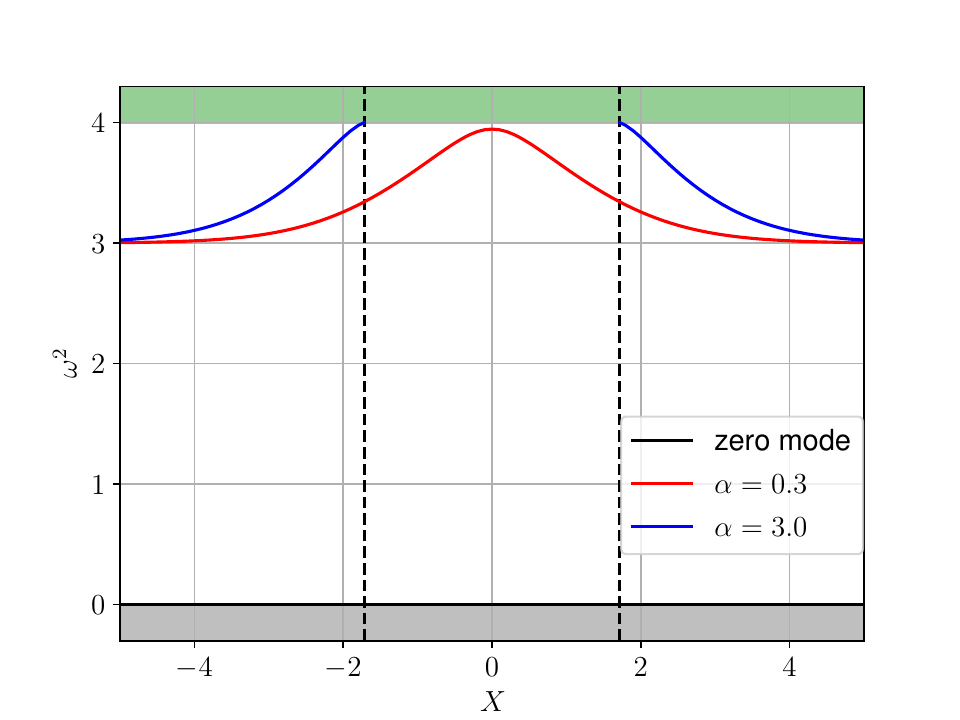}
    \caption{Kink's bosonic and fermionic spectrum as a function of $X$.}
    \label{fig:spec}
\end{figure}
We computed the kink's spectrum as a function of the parameter $X$. This is shown in Fig.~\ref{fig:spec}. For large $|X|$, the excited state frequency $\omega$
 coincides with the frequency of the shape mode in the absence of impurity, as expected. However, it increases near $X=0$. If $\alpha$ is large 
enough, $\omega$ reaches the threshold value, the junction between the bound and continuum spectrum. This happens at two critical points, symmetric
 around $X=0$. For $\alpha=3.0$, we find the critical values to be $X_{sw}=\pm1.718$.
 The points at which a mode hits the mass threshold will be further linked to the appearance of spectral walls.
   
In order to obtain the fermionic modes, we consider the following ansatz:
\begin{equation}
\psi(t,x)=e^{-i \epsilon_r t}\psi(x).
\end{equation}
As a result, the fermionic modes satisfy
\begin{eqnarray}
\label{lag:fer-11}
i\partial_x \psi^r_2+i \mathcal{U}\psi^r_2&=&\epsilon_r \psi_1^r,\\\label{lag:fer-12}
i\partial_x \psi^r_1-i \mathcal{U} \psi^r_1&=&\epsilon_r \psi^r_2.
\end{eqnarray}
Then the second-order decoupled equations for up and down components take the form:
\begin{eqnarray}
\left(-\partial_x^2 + V_-\right)\psi^r_1&=&\epsilon_r^2\psi^r_1 \label{fer:eigen_11},\\
\left(-\partial_x^2 +V_+\right)\psi^r_2&=&\epsilon_r^2\psi^r_2 \label{fer:eigen_21}.
\end{eqnarray}

Let us note that the equations (\ref{bos:eigen}) and (\ref{fer:eigen_11}) are the same. This comes from SUSY, guaranteeing the same spectra for
 the boson and fermion fields.

In the language of SUSY quantum mechanics, the potentials $V_{\pm}$ in (\ref{fer:eigen_11}) and (\ref{fer:eigen_21}) are supersymmetric partners. 
For $\epsilon_0=0$, their solutions correspond to the fermionic zero mode and coincide with the solutions (\ref{fer-zero}).
Using standard results, it is trivial to show that, except for the zero mode (where $\psi_2^r$ is trivial), both equations have identical spectra. 
 Therefore, for a given $(\alpha, X)$, if (\ref{bos:eigen}) has $n$ normalizable modes, then (\ref{fer:eigen_11}) has $n$ normalizable modes
 and (\ref{fer:eigen_21}) has $(n-1)$ plus a trivial one.
 Consequently, the number of fermionic and bosonic modes is the same. 

Obviously, in a complete analysis, one should also consider the coupled linear perturbation problem. Depending on the value of the back-reaction
 parameter, this would lead to some modifications in the linear spectrum. Fortunately, for our purposes, it is enough to stay in the decoupled regime. 

Let us also underline that the fermionic mode has to have a uniquely defined amplitude due to the normalization condition. Contrary to the bosonic shape mode, its amplitude is not a free parameter of the initial data.

\section{Spectral wall in the presence of fermions}

\subsection{Bosonic spectral wall}
Before studying how the fermion field affects the spectral wall phenomenon, let us briefly describe the spectral wall phenomenon in the bosonic
 sector with $\psi=0$ \cite{Adam:2019xuc}.
 To do this, we integrate the equation of motion, which is given by
\begin{equation}
\phi_{tt}=\phi_{xx}+2 (\phi-\phi^3)(1+\sigma)^2+\sigma_x(1-\phi^2).\\
\end{equation}
For the initial condition we take
\begin{equation}
\label{sol-shape}
\phi(x,t)=-\tanh\left[\gamma(x+X_0-vt)\right]+A\cos(\sqrt{3}t)\tanh\left[\gamma(x+X_0-vt)\right]\sech\left[\gamma(x+X_0-vt)\right],
\end{equation}
which describes an antikink far from the impurity with the excited shape mode. The parameter $A$ describes the amplitude of the wobbling. 
If the shape mode is not excited, the evolution goes along the sequence of the energetically equivalent BPS configurations. This can be very 
accurately described by a geodesic motion arising from a collective 
model based on the BPS solutions with the modulus $c$ \cite{manton2004topological}. 

Next, we assume that the shape mode is excited, carrying some part of the energy. As the antikink approaches the impurity, the frequency of the
 mode rises, and finally,  at $X=X_{sw}$, it equals the mass threshold. At this point, the normal mode becomes a non-normalizable threshold mode.
 This has a profound effect on the soliton dynamics.

To see this, we consider the antikink whose initial position is $X=10.0$, which is far enough so that the impurity does not deform the antikink.
Let us take the initial velocity of the antikink as $v=0.01$. The equations can be integrated in a box in the interval $-100.0<x<100.0$ with
 anti-periodic boundary conditions, divided by step size $\Delta x=0.05$. The space derivatives are computed using a five-point stencil approximation. 
The resulting set of equations has been integrated using a fifth-order Runge-Kutta method with adaptive step size and error control. The energy conservation has been measured throughout the evolution, and the maximum error is of order $10^{-5}$.

\begin{figure}
    \centering
    \includegraphics[width=0.8\textwidth]{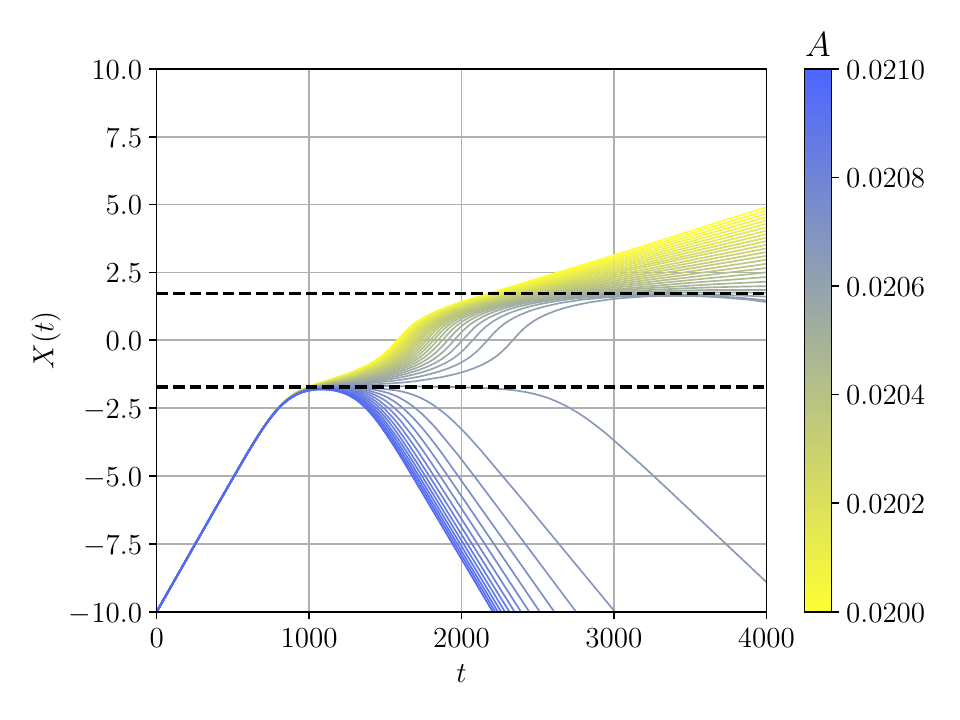}
    \caption{Antikink's position as a function of time for several values of the shape mode amplitude.}
    \label{fig:center-boson}
\end{figure}
The path taken by the antikink is shown in Fig.~\ref{fig:center-boson} for several values of $A$. The behavior strongly depends on 
the amplitude of the mode.
 For small amplitudes, $A<A_{cr}$, the antikink passes through the point where the normal mode disappears $x=X_{sw}$. The distortion of the
 geodesic motion becomes smaller as the amplitude decreases. For large amplitudes, $A>A_{cr}$, the antikink is back-scattered before $x=X_{sw}$.
 The reflection occurs earlier for larger values of the amplitude. 
Finally, for the critical amplitude  $A_{cr}\simeq0.0207$, the antikink forms a quasi-stationary state located precisely at $x=X_{sw}$. 
This is the position of the spectral wall. This value is independent of the initial velocity. 

\subsection{Spectral wall with fermions}

Next, we consider the full supersymmetric model. A natural generalization of the collision in the purely bosonic sector is to choose for the
 initial configuration an excited antikink, carrying the shape mode, bound with an unexcited fermion field. This configuration is then boosted
 towards impurity. Thus, our initial configuration at $t=0$ is the same as (\ref{sol-shape}) and
\begin{equation}
\psi(x,t=0)=N \Lambda
\begin{pmatrix}
 1/\cosh^2[\gamma(x+X_0)]  \\
0
\end{pmatrix},
\end{equation}
where
\begin{equation}
\Lambda=\begin{pmatrix}
\cosh(\chi/2)&-\sinh(\chi/2)\\
-\sinh(\chi/2)&\cosh(\chi/2)
\end{pmatrix},
\end{equation}
with $\chi=\tanh^{-1}(v)$ and $N$ chosen to fulfill the normalization condition. The matrix $\Lambda$ is the boost operator. The fermion contribution is just a boosted zero mode, and 
therefore $\bar{\psi} \psi |_{t=0} 
= i[\psi_2^*(x,t)\psi_1(x,t)-\psi_1^*(x,t)\psi_2(x,t)]|_{t=0} =0$. In other words, for this initial state, there is no back-reaction
 of the fermionic field in the bosonic equation of motion. As it was shown in \cite{campos2022kink}, this implies that $\bar{\psi} \psi$ term 
vanishes for any $t>0$. Therefore, there is no back reaction during the whole evolution. Thus, the dynamics of the kink is {\it exactly} as in
 the purely bosonic version of the model, and the spectral wall remains intact. 
Furthermore, the fermion at the zero mode stays bound to the antikink at any time $t>0$.

To allow for a non-zero back reaction, we need to consider an initial condition that goes beyond the fermionic zero mode. Here, an obvious choice is the first excited fermion state. 
The initial condition in our simulations now takes the form
\begin{equation}
\phi(x,t)=-\tanh\left[\gamma(x+X_0-vt)\right],
\end{equation}
and
\begin{equation}
\psi(x,t=0)=N \Lambda\begin{pmatrix}
\sqrt{3}\tanh\left[\gamma(x+X_0)\right]\sech\left[\gamma(x+X_0)\right]\\
i\sech\left[\gamma(x+X_0)\right]
\end{pmatrix},
\end{equation}
The amplitude $N$ is again fixed by the normalization of the fermion field. The scalar field corresponds to an unexcited antikink, while the fermion field corresponds to the solution for the fermion's first excited state. Importantly, the fermion field is entirely in the excited state.  
The parameters are again $v=0.01$ 
and $X_0=10.0$. Moreover, the employed numerical methods are the same as before, and we obtain a maximum error in the total energy of the order of $10^{-4}$. To confirm our numerical results, we also varied the box size. 
\begin{figure}
    \centering
    \includegraphics[width=0.8\textwidth]{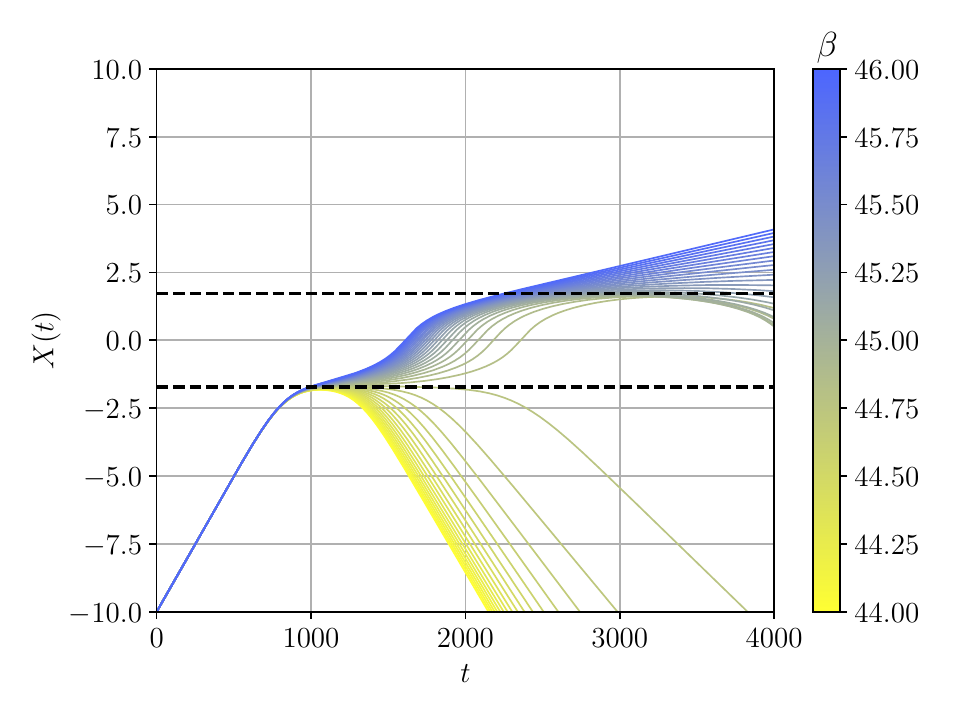}
    \caption{Antikink's position as a function of time for several values of the back-reaction coupling constant $\beta$.}
    \label{fig:center-fermion}
\end{figure}
\begin{figure}
    \centering
    \includegraphics[width=\textwidth]{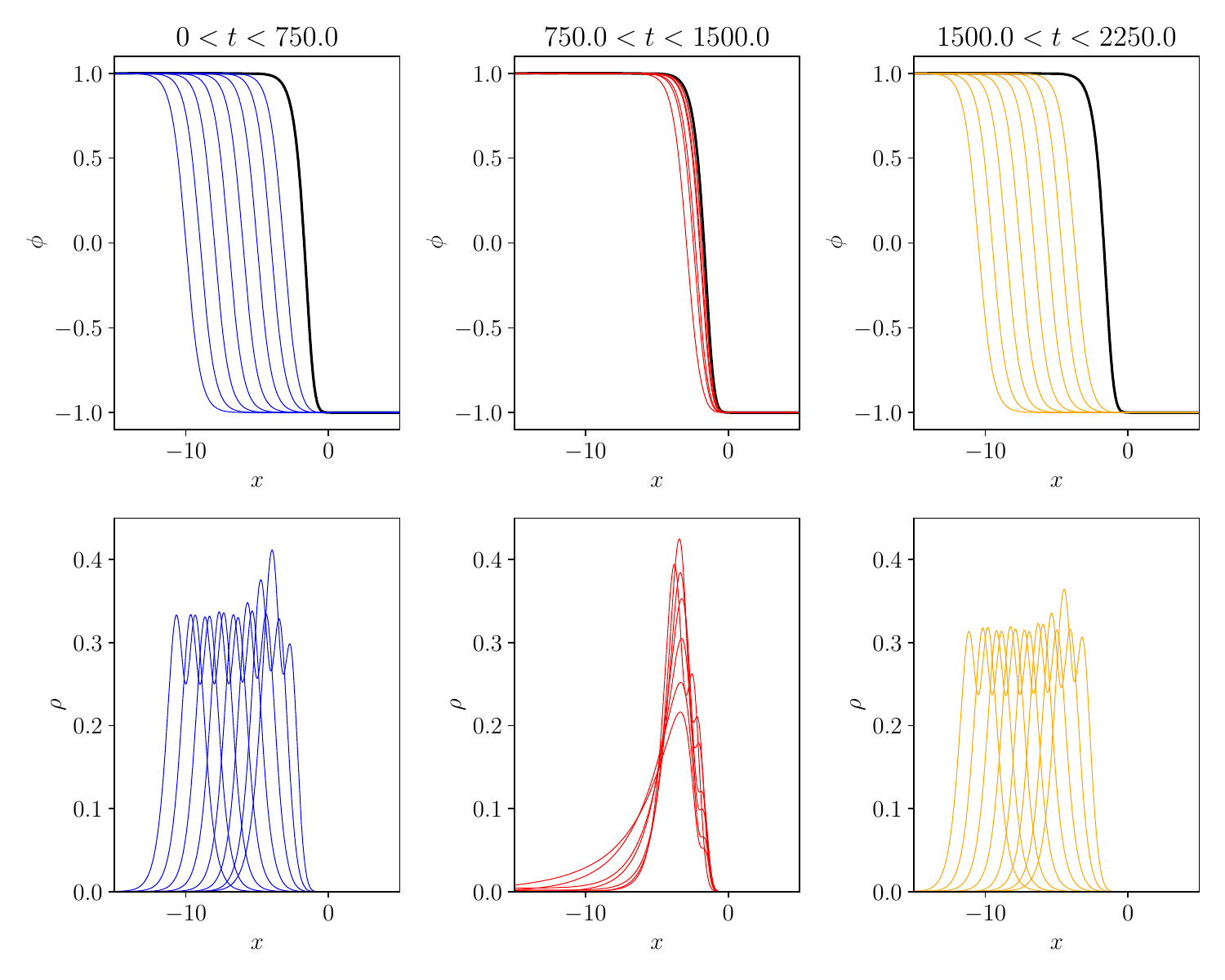}
    \caption{Several snapshots of the scalar field $\phi$ and fermion number density $\rho$ as the antikink approaches the spectral wall. 
The antikink is reflected by the wall for $\beta=44.0$.}
    \label{fig:snap44}
\end{figure}

The results of our simulations are shown in Fig.~\ref{fig:center-fermion}. Interestingly, we also find a spectral wall located exactly in 
the correct position. Now, this is triggered by the fermion excited state. In the no-back reaction spectral problem approximation, this state
 hits the mass threshold precisely at the same $X=X_{sw}$. The fact that we see the wall at this position confirms the correctness of this
 approximation for the spectral problem. We remark that, due to the fixed normalization of the initial fermion field,  the only parameter 
that can be varied is the strength of the back-reaction $\beta$. This is a rather fundamental difference to the bosonic sector where the amplitude of the mode is an adjustable initial parameter. In our simulations, we took $\beta \in [44,46]$. We find that the quasi-stationary solution is formed when this coupling takes the following critical value $\beta_{cr} \simeq 44.8$.

Due to the coupling between bosonic and fermionic sectors, it could be that the fermion field excites the bosonic field, and the 
spectral wall comes from this bosonic excitation. To verify that, we subtracted the kink solution from the scalar field and projected the 
result into the bosonic shape mode. We found that the amplitude of the projection is much smaller than the necessary amplitude to create a 
bosonic spectral wall. Hence, it is indeed the fermion mode responsible for the appearance of the spectral wall. 

\begin{figure}
    \centering
    \includegraphics[width=\textwidth]{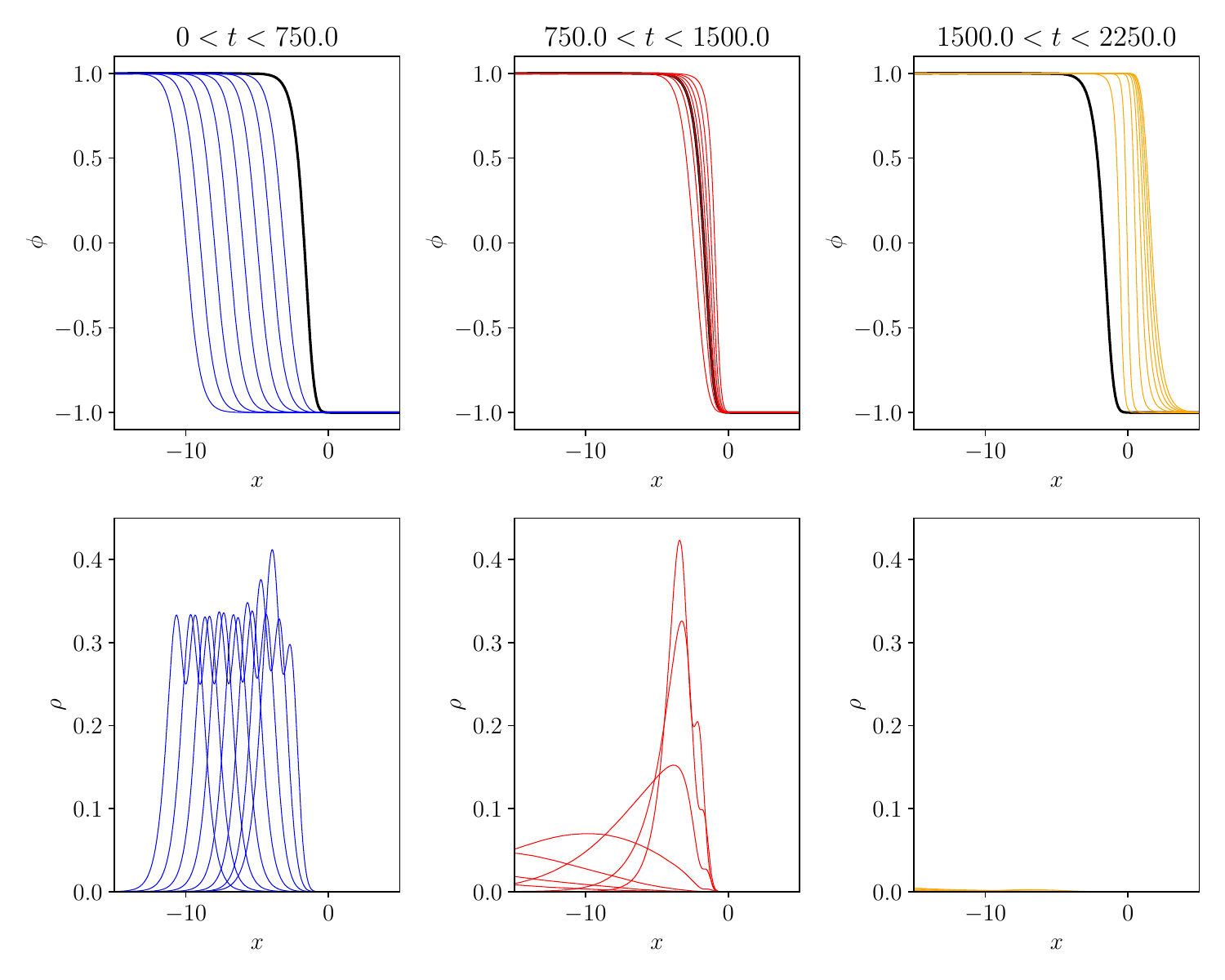}
    \caption{Several snapshots of the scalar field $\phi$ and fermion number density $\rho$ as the antikink approaches the spectral wall.
 The antikink crosses the wall for $\beta=46.0$.}
    \label{fig:snap46}
\end{figure}

To better understand the behavior of the antikink as it approaches the spectral wall, we show snapshots of its evolution in Figs.~\ref{fig:snap44} 
and \ref{fig:snap46}. They show the evolution right below the critical value of the parameter and right above it, respectively. In Fig.~\ref{fig:snap44}, we 
consider $\beta=44.0$. In this case, the antikink is reflected by the wall, marked in black. The left panels show the antikink approaching the wall. 
In such a case, the fermion number density $\rho=\psi^\dagger\psi$ is localized and has two peaks, as expected for the first excited state. One can see 
that closer to the wall (and the impurity), $\rho$ gets slightly deformed. Then, in the middle panels, we see that the antikink gets temporarily trapped at the wall.
Meanwhile, the fermion is still bound to the kink, becoming more delocalized at the closest point to the wall. As it stays attached to the antikink,
 both reflect from the wall, as shown in the right panels. The kink localized part of the fermion field recovers the same shape after the interaction,
 although with a smaller amplitude due to radiative losses. This means that a part of the fermionic field transits into continuum states. 

\begin{figure}
    \centering
  \includegraphics[width=0.8\textwidth]{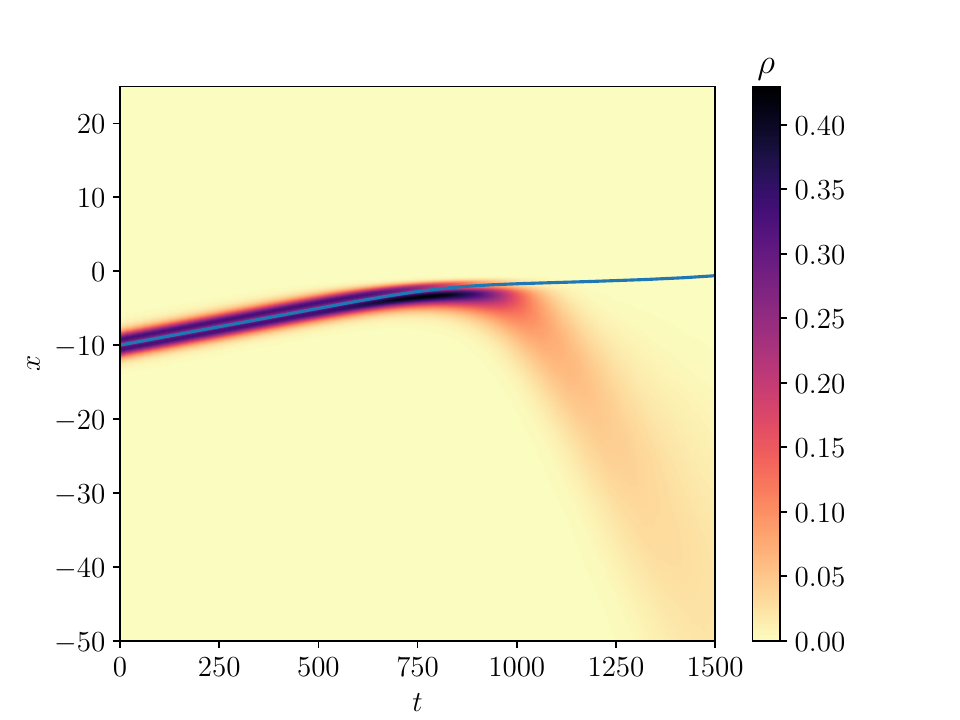}
    \caption{Evolution of the fermion number density in spacetime as the antikink crosses the spectral wall. The fermion is reflected by the wall 
and decouples from the kink. The positive energy fermion is in the first excited state. $\beta=46.0$.}
    \label{fig:field46}
\end{figure}

A more surprising scenario occurs for $\beta=46.0$, which is above the critical value, implying that the kink crosses the wall. Corresponding snapshots are 
shown in Fig.~\ref{fig:snap46}. In the left panels, the antikink approaches the wall, and we observe the same behavior as before. In the middle panels,
 we see that the antikink is temporarily trapped at the wall but then crosses it. A somewhat different behavior reveals the fermionic field, which cannot pass through the spectral wall. Indeed, the excited fermionic mode does not exist beyond this point. Therefore the fermion separates from its bosonic partner. Once it happens, the energy stored in the excited fermionic bound mode is no longer stable. Such a fermionic state becomes delocalized and quickly decays into radiation which moves in the direction opposite to the free antikink. The reason is that the created fermionic radiation possesses low energy with a small wave number to overcome the potential barrier arising from the impurity. As the motion continues, we see in the right panels that the antikink continues moving to the right, while the fermion is not observed in the
 same range. The evolution of $\rho$ can explain the actual fermion behavior in spacetime during the process. The fermion number density is shown in 
Fig.~\ref{fig:field46}. The solid blue line shows the trajectory of the antikink center. One can observe that as the antikink moves past the wall,
 the fermion unbinds from the kink and gets reflected. This implies that the fermion strongly feels the presence of the wall.

We can draw a fascinating picture of the spectral wall phenomenon with fermions. On one hand, the fermion is not allowed to
move through it. This is because of the fact that initially, it is entirely in the excited state, which ceases to exist after $x>X_{sw}$. Therefore, irrespectively of the strength of the back reaction, such a fermion is never allowed to cross the spectral wall and is always backscattered. On the other hand, the bosonic field, i.e., antikink, is now blind to the wall as the relevant mode is basically unexcited (or excited dynamically during the collision in a tiny amount). Hence, the antikink tends to pass the wall. However, it is the boson-fermion coupling terms that may prevent the antikink from passing the spectral wall. Indeed, if the back-reaction is strong, the fermion field pulls back the antikink. If the back reaction is weak, it cannot do so, and
 the antikink continues its trajectory. In conclusion, the spectral wall behaves as a filter for excited fermion-kink bound states.

\begin{figure}
    \centering
        \includegraphics[width=0.8\textwidth]{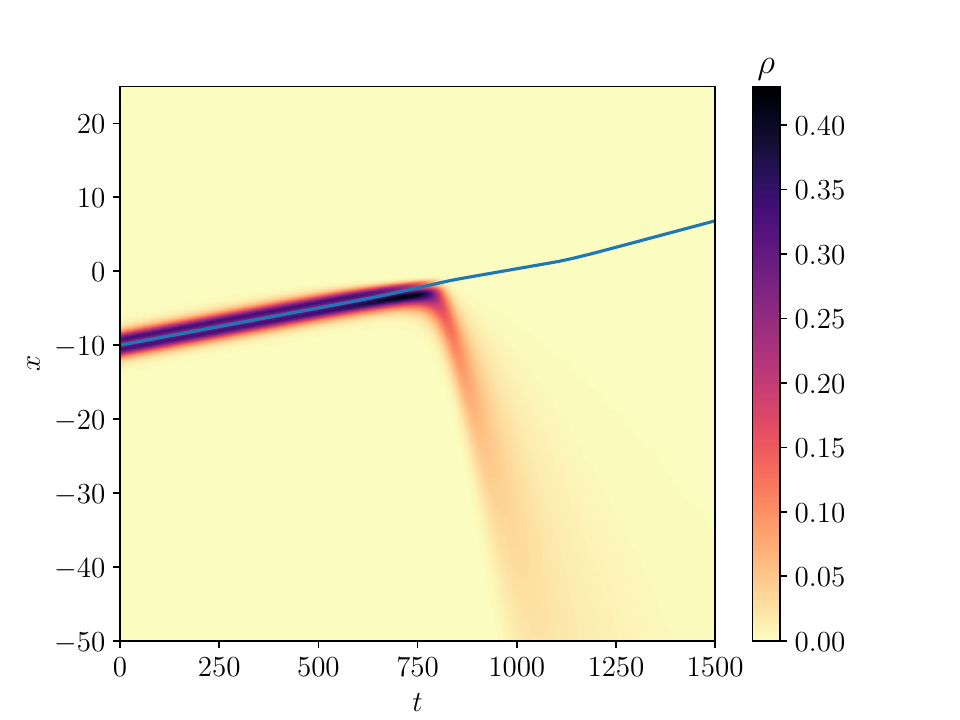}
    \caption{Evolution of the fermion number density in spacetime as the antikink crosses the spectral wall. The fermion is reflected by the wall 
and decouples from the kink. The negative energy fermion is in the first excited state. $\beta=46.0$.}
    \label{fig:fieldneg}
\end{figure}

The same scenario repeats when the fermion is initially in the first excited state but with negative energy. The antikink can pass the spectral wall while the fermionic field is stopped and finally spreads via radiation, see Fig. \ref{fig:fieldneg}.

\begin{figure}
    \centering
    \includegraphics[width=0.8\textwidth]{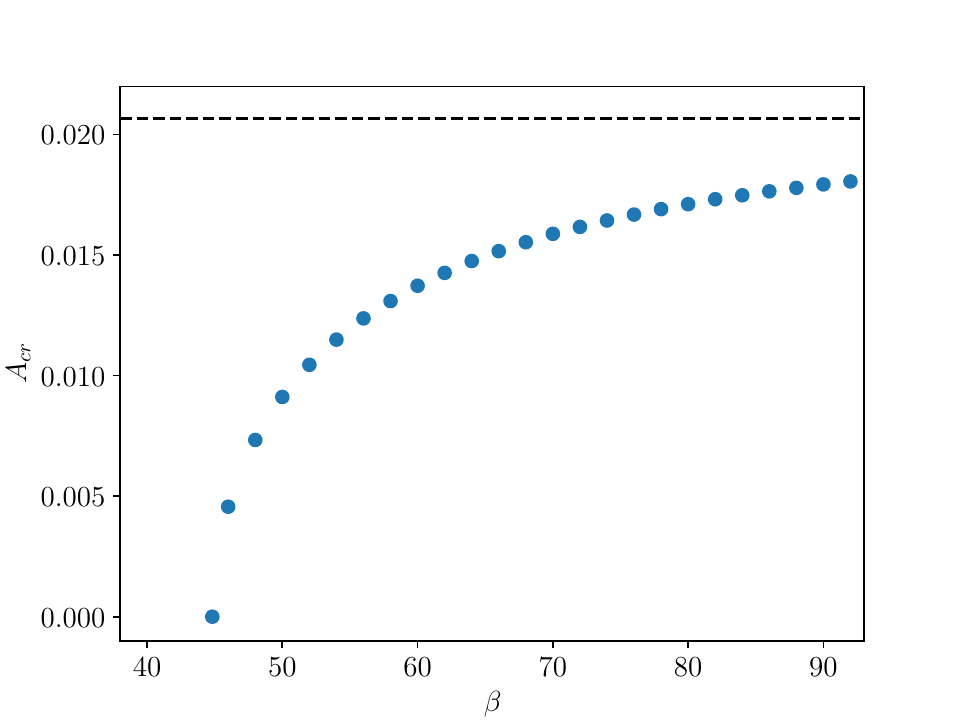}
    \caption{Critical amplitude of the shape mode as a function of the back-reaction coupling constant $\beta$. }
    \label{fig:crit}
\end{figure}

Finally, we considered the initial state where both boson and fermion fields carry some excitations. Specifically in the boson-fermion bound state, 
  initially, the bosonic field is the antikink with the shape mode excited, while the fermionic part consists of the first excited state. Once again, we underline that only the amplitude of the bosonic mode is subject to change. The normalization condition fixes the strength of the fermionic mode. Now, the spectral wall can occur by combining a fermion excitation and shape mode excitation. The impact of the fermionic mode is visible in a change of the value of the critical amplitude of the shape mode at which we observe the formation of the quasi-stationary state. It is, as expected, all the time located at $x=X_{sw}$, that is the point at which both modes enter the continuum spectrum. Now, the critical amplitude is a function of the coupling $\beta$. This is illustrated in Fig.~\ref{fig:crit}. The interpretation is again in full accordance with the usual purely bosonic spectral wall phenomenon.

For theories with a weaker back-reaction, $\beta > \beta_{cr}$, there is a critical value of the shape mode amplitude $A=A_{cr}(\beta)$ for which the quasi-stationary state is formed. For $A<A_{cr}(\beta)$, the kink passes the spectral wall, while for $A>A_{cr}(\beta)$, the kink-fermion incoming state is backscattered. We remark that for all values of $\beta$, the value of the critical amplitude is smaller than the value of the critical amplitude for the decoupled case, $A_{cr}(\beta)< A_{cr}(\infty)$. The interpretation is straightforward. The excited fermion contributes to the formation of the stationary state (or in general to the spectral wall phenomenon) together with the bosonic shape mode.



\section{Conclusions}

In the present work, we have established the existence of spectral walls in a supersymmetric model. Their main properties, like the 
independence of the position of a spectral wall for all initial conditions or their selective nature, remain basically unchanged. However, due
 to a richer field content, some novel effects additionally emerge. 

Firstly, a kink-fermion bound state can experience a spectral wall even though {\it only} the fermionic field is excited. 
Interestingly, when a fermion bound mode enters the continuum, the fermion feels a spectral wall as it happens for the bosonic case with wobbling kinks. This way, the fermion bound mode plays the same role as the ``shape mode" for the kink. In the supersymmetric model,
the excited fermion state ceases to exist precisely at the same point as the bosonic normal mode. Here, we once again underline an important difference in the nature of the bosonic and fermionic bound modes. While the bosonic mode, like e.g., the shape mode, can be excited with an arbitrary amplitude, the amplitude of the fermionic mode is uniquely determined by the normalization condition imposed on the fermion field. Thus, in a given theory, that is for a given value of the back reaction coupling constant $\beta$, only one scenario may occur. Namely, for the critical value of the coupling, $\beta=\beta_{cr}$, the antikink forms a long-living quasi-stationary state whose position $x=X_{sw}$ is governed by the point where the fermionic mode hits the mass threshold. For weaker coupling, $\beta>\beta_{cr}$, the antikink can pass through the spectral wall, while for stronger coupling, $\beta<\beta_{cr}$, it is backscattered before $x=X_{sw}$. 

In the case when initially also the bosonic shape mode of the antikink is excited, the inclusion of fermionic bound mode modifies the value of the critical amplitude of the shape mode $A_{cr}$ at which the stationary state is formed. This value depends on the coupling $\beta$. 

Secondly, what is more captivating, a spectral wall may act not only as a barrier separating more excited from less excited kink-fermion bound states but also may separate the bosonic from fermionic fields leading to a disintegration of the bound
 state. Strictly speaking, we observed that the initially excited  
fermionic mode is always reflected at the spectral wall, while the bosonic antikink may or may not pass through it. The strength of the back reaction determines the actual behavior at the spectral wall. If the back reaction is not too strong, the initially bounded kink-fermion state is destroyed, and the bosonic and fermionic degrees of freedom evolve in different directions. Otherwise, if the back-reaction is sufficiently strong, the fermion stays bound to the bosonic field and pulls it back with itself. As a result, the whole kink-fermion bound state is backscattered by the spectral wall.

The observed ability to disintegrate a bound state seems to be a rather generic property of spectral walls, which should occur in other theories, e.g., two-scalar field models, provided there is a bound state with a finite, or preferably small, binding energy. It would be interesting to present an explicit example of such a process. This could be investigated using results of \cite{HB} in theories for which stable and meta-stable multikink states have been constructed. 

It is possible that this effect may also have an impact on the semi-classical quantization of solitons with small binding energies. Indeed, such quantum corrections are based mainly on zero, and massive bound, or vibrational modes \cite{Hal-1, Hal-2, Gud}. Whether a (quantum) spectral wall may lead to the instability of weakly bound solitons requires independent studies. Here the framework developed in \cite{Jarah-1} can be very useful; see also \cite{Jarah-2} for its recent application to spectral walls.  
\section*{Acknowledgements}
J. G. F. C. and A. M. acknowledge financial support from the National Council for Scientific and Technological Development - CNPq, Grant nos. 150166/2022-2 and 309368/2020-0, respectively. A. M. is also supported by the Brazilian agency CAPES and Universidade Federal de Pernambuco Edital Qualis A. J. M. Q. is supported by the Spanish Ministry of Science and Innovation (Project No. PID2020-113406GB-I00). A. W. was supported by the Polish National Science Centre (Grant No. NCN 2019/35/B/ST2/00059). Part of the simulations performed in the current work was done in the Brazilian supercomputer SDumont from the Laboratório Nacional de Computação Científica.


\end{document}